\documentclass[preprint,pteplogo]{ptephy_v1}

\preprintnumber{XXXX-XXXX} %%% %%% Insert preprint number here

\usepackage{amssymb}
\usepackage{amsmath}
\usepackage{graphicx}
\usepackage{color}
\usepackage{slashed}
\newcommand{\be}{\begin{equation}}
\newcommand{\ee}{\end{equation}}
\newcommand{\ben}{\begin{eqnarray}}
\newcommand{\een}{\end{eqnarray}}

\begin{document}

\title{Analysis of $X(3872)$ production via Heavy-Meson Effective Theory }
%\shorttitle{Analysis of $X(3872)$ production via Heavy-Meson Effective Theory }

\author{L. M. Abreu}
%\shortauthor{L. M. Abreu}

\affil{Instituto de F{\'i}sica, Universidade Federal da Bahia, 40210-340, Salvador, BA, Brazil \email{luciano.abreu@ufba.br}}

%\ead{luciano.abreu@ufba.br}
%\pacs{12.39.Hg}{Heavy quark effective theory}
%\pacs{12.39.Mk}{Glueball and nonstandard multi-quark/gluon states}
%\pacs{14.40.Rt}{Exotic mesons}
%\pacs{13.75.Lb}{Meson-meson interactions}

%\vspace{10pt}

\begin{abstract}
  We analyze the $X(3872)$ production in the processes $ \bar{D} D \rightarrow \pi X $, $\bar{D}^* D \rightarrow \pi X $ and $\bar{D}^* D^* \rightarrow \pi X $, making use of the Heavy-Meson Effective Theory, with an effective Lagrangian built based on chiral $SU(3)_{L} \times SU(3)_{R}$ and heavy quark symmetries. In this scenario, we consider the $X(3872)$ as a bound state of $D^{*} \bar{D} + c.c.$, including neutral and charged components, and obtain the cross sections the mentioned reactions as function of collision energy.
\end{abstract} 

%\pacs{12.39.Hg, 14.40.Rt, 12.39.Mk, 13.75.Lb}

%\keywords{heavy-meson effective theory, meson-meson bound states, exotic hadron states}

\subjectindex{B36, B60, B69, B37, D32}

\maketitle

%%%%%%%%%%%%%%%%%%%%%%%%%%%%%%%%%%%%%%%%%%%%%%%%%%%%%%%%%%%%%%%%%%%%
%%%%%%%%%%%%%%%%%%%%%%%%%%%%%%%%%%%%%%%%%%%%%%%%%%%%%%%%%%%%%%%%%%%%
%%%%%%%%%%%%%%%%%%%%%%%%%%%%%%%%%%%%%%%%%%%%%%%%%%%%%%%%%%%%%%%%%%%%

\section{Introduction}
%%%%%%%%%%%%%%%%%%%%%%%%%%%%%%%%%%%%%%%%%%%%%%%%%%%%%%%%%%%%%%%%%%%%
%%%%%%%%%%%%%%%%%%%%%%%%%%%%%%%%%%%%%%%%%%%%%%%%%%%%%%%%%%%%%%%%%%%%
%%%%%%%%%%%%%%%%%%%%%%%%%%%%%%%%%%%%%%%%%%%%%%%%%%%%%%%%%%%%%%%%%%%%

Observations of unconventional hadron states in recent years have been reported by several experiments, both in charmonium and bottomonium spectra \cite{Brambilla,PDG}. From a theoretical point of view, a large amount of effort has been directed to understand their structure, and
several interpretations and models have been proposed~\cite{Brambilla2,Chen}. 

One emblematic example is the state $X(3872)$, first discovered in 2003 by  Belle Collaboration \cite{Choi:2003ue}, and afterward  
detected by other facilities \cite{Choi:2003ue,Aubert:2008gu,Acosta:2003zx,Abulencia:2006ma,Aaltonen:2009vj,Abazov:2004kp,Aaij:2011sn,Aaij:2013zoa,Chatrchyan:2013cld}. Its quantum numbers have been established recently: $J^{PC}=1^{++}$~\cite{Aaij:2013zoa}. Concerning its structure, several proposals have been attempted in order to provide some insight \cite{Brambilla2,Chen}. We remark three interpretations of $X(3872)$ state that have been largely explored: the tetraquark state picture, i.e. a binding of a diquark and an antidiquark \cite{Maiani:2004vq,Esposito1};  the radial excitation of the axial vector charmonium state \cite{Fazio,Badalian,Ortega,Ferretti1,Ferretti2,Karliner,Takizawa}; and a loosely bound state of $(D^{*0}\bar{D}^0+c.c.)$~\cite{Tornqvist3,AlFiky,Braaten1,Dong1,Dong2,Liu,Lee,Braaten2,Nieves,Aceti,Hidalgo,Guo}.

Motivated to get useful physical quantities in the determination of the $X(3872)$ structure, the authors of Refs.~\cite{ChoLee,XProduction,Artoisenet,Esposito2013,Guerrieri} have performed investigations about the relevance of light mesons for the production of $X(3872)$. In particular, Ref.~\cite{ChoLee} studied the 
hadronic effects on $X(3872)$ abundance, showing the necessity of evaluation of the interaction among $ X(3872)$ and light hadrons, since it can be absorbed by the comoving  light mesons or produced from the interaction between charmed mesons (i.e. reactions like $D^{(*)} \bar{D}^{(*)} \to X \pi$). The amplitudes have been determined by taking into account the neutral components of $X$ and effective Lagrangians furnishing $D \bar{D}^{\ast} \pi$ and  $D D^{\ast} X$ vertices. 
Besides, it is also worthy of mention the results obtained in Ref.~\cite{XProduction}: based on a $SU(4)$ effective approach, with the incorporation of anomalous vertices $D^*\bar{D}^* \pi$ and $ D^*\bar{D}^* X$, and also the adding of the charged components of the $D$ and
$D^*$ mesons in the coupling to $X$, the $D^{(*)}\bar{D}^{(*)}\to X+\pi$ cross-sections try out a relevant enhancement. 

Thus, inspired by the findings of Refs.~\cite{ChoLee,XProduction}, 
in this work we analyze the processes $ \bar{D} D \rightarrow \pi X $, $\bar{D}^* D \rightarrow \pi X $ and $\bar{D}^* D^* \rightarrow \pi X $ in another perspective: we work within the framework of Heavy-Meson Effective Theory (HMET), with an effective Lagrangian built respecting chiral $SU(3)_{L} \times SU(3)_{R}$ and heavy quark symmetries.  We consider $X(3872)$ state as a bound state of $(D^* \bar{D} + c.c.)$, including neutral and charged components, and obtain the cross sections of the mentioned reactions.

This work is organized as follows.  In Section~\ref{Formalism}, we present the 
formalism of HMET and obtain the interaction Lagrangian with the relevant vertices. 
Following, in Section~\ref{Amplitudes} the transition amplitudes and cross sections are determined and analyzed. 
After that, we discuss in Section ~\ref{Discussion} our results and compare our approach to other ones available in literature. 
We summarize the results and conclusions in Section~\ref{Conclusion}.  The Feynman Rules of the theory are presented in Appendix \ref{AppA}.

%%%%%%%%%%%%%%%%%%%%%%%%%%%%%%%%%%%%%%%%%%%%%%%%%%%%%%%%%%%%%%%%%%%%
%%%%%%%%%%%%%%%%%%%%%%%%%%%%%%%%%%%%%%%%%%%%%%%%%%%%%%%%%%%%%%%%%%%%
%%%%%%%%%%%%%%%%%%%%%%%%%%%%%%%%%%%%%%%%%%%%%%%%%%%%%%%%%%%%%%%%%%%%

\section{Formalism}
\label{Formalism}
%%%%%%%%%%%%%%%%%%%%%%%%%%%%%%%%%%%%%%%%%%%%%%%%%%%%%%%%%%%%%%%%%%%%
%%%%%%%%%%%%%%%%%%%%%%%%%%%%%%%%%%%%%%%%%%%%%%%%%%%%%%%%%%%%%%%%%%%%
%%%%%%%%%%%%%%%%%%%%%%%%%%%%%%%%%%%%%%%%%%%%%%%%%%%%%%%%%%%%%%%%%%%%

In order to investigate the processes involving the $X(3872)$ production, we must consider an effective theory that describes the interactions 
between heavy mesons, i.e. mesons containing a heavy quark $Q$. 
Thus, we work under the effective theory known as Heavy Meson Effective Theory (HMET)
 \cite{Ingsur,Eichten,Georgi,Grinstein,Manohar,Casalbuoni,Abreu,Esposito:2014rxa}. 
On this subject, we define the superfields: 
\begin{eqnarray}
H_a ^{(Q)} & = & \left( \frac{1+ v_{\mu} \gamma ^{\mu} }{2}\right)\left( 
P _{a \mu}^{* (Q)} 
\gamma ^{\mu}  - P _{a }^{ (Q)}  \gamma ^{5} \right) ,\nonumber \\
H ^{(\bar{Q}) a} & = & \left( P _{ \mu}^{* (\bar{Q}) a} \gamma ^{\mu} - 
P ^{ (\bar{Q}) a}  \gamma^{5} \right) \left( \frac{1 - v_{\mu} 
\gamma ^{\mu}}{2}\right) .
\label{H2}
\end{eqnarray}
In Eq. (\ref{H2}), $Q = c,b$ is the index with respect to the heavy-quark flavor group $SU(2)_{HF}$; $v$ is the velocity parameter; $a$ is the triplet index of 
the $SU(3)_V$ group; and
$P _{a }^{ (Q/ \bar{Q})}$ and $P _{a \mu}^{* (Q / \bar{Q})} $ are
the pseudoscalar and vector heavy-meson fields forming a $\mathbf{\bar{3}}$
representation of $SU(3)_V$, i.e.
\begin{eqnarray}
  P _{a }^{ (c)} & = & \left( D^{0},  D^{+}, D_s ^{+} \right), \nonumber \\
  P _{a }^{ (\bar{c})} & = & \left( \bar{D}^{0},  D^{-}, D_s ^{-} \right) ,
  \label{P1}
\end{eqnarray}
for the charmed pseudoscalar meson field, and analogous expressions for the vector case. 
Notice that the heavy vector meson fields obey the conditions: 
\begin{eqnarray}
 v \cdot  P _{a }^{* (Q)} & = & 0, \nonumber \\
  v \cdot  P ^{* (\bar{Q}) a} & = & 0 . 
\label{C1}  
 \end{eqnarray}
They define the three different polarizations of the heavy vector mesons.

Focusing on the $X(3872)$ state, we assume it as an elementary degree of freedom of our formalism with quantum numbers $J^{PC}= 1^{++}$. Since we work within HMET whose guiding principle is Heavy-Quark Spin Symmetry, it seems natural to construct the $X$-field accordingly, i.e. associating it to a superfield with similar quantum numbers. In this sense, inspired by the discussion presented in Refs. \cite{Fazio,Casalbuoni,Esposito:2014rxa} of covariant fields for $S$-wave and $P$-wave states, we assume that $X(3872)$ can be represented by the superfield
\be  
\mathcal{X} ^{\mu} = \left( \frac{1+ v_{\rho} \gamma ^{\rho} }{2}\right)
\left[ \frac{1}{2} \epsilon ^{\mu \alpha \beta \gamma} v_{\alpha } \gamma _{\beta}
X_{\gamma} \right]
\left( \frac{1 - v_{\sigma} 
\gamma ^{\sigma}}{2}\right) ,
\label{Xfield}
\ee
where $X$ is the quantized field associated to $X(3872)$ state. 

We stress that the superfield $\mathcal{X} ^{\mu}$ defined in Eq. (\ref{Xfield}) has only the $J^{PC}= 1^{++}$ component, since no spin partner for the $X(3872)$ is involved. Notwithstanding, it is relevant noticing that attempts to associate superfields with exotic fields in HMET, even without spin partners, have been performed in several works. 
For example, in Ref. \cite{Esposito:2014rxa} the exotic states $Z_c(3900)^{+}$ and $Z_c '(4020)$ are considered as a superfield with only one component (without spin partners) associated to known charmonium field with similar quantum numbers, giving reasonable results. Besides, an equivalent procedure for superfields representing $Z_b ^{(\prime)}$ states has been done in Ref.~\cite{Cleven}. Coming back to the $X(3872)$ case, in Ref.~\cite{Fazio} the $X$ state has been defined  in terms of superfield, in the scenario of radially excited charmonium picture.

Thus, in our understanding the analysis and description of properties of exotic hadronic states via the HMET approach stand on strong ground, and a consistent and more rigorous theory can be constructed by invoking Heavy-Quark Spin Symmetry as fundamental concept. In this sense, as it will be shown later the definition in Eq. (\ref{Xfield}) is useful for construction of an effective Lagrangian consistent with other versions discussed in different papers (see Refs. \cite{Dong2,Liu,Aceti,XProduction}).

To construct invariant quantities under relevant symmetries, we need 
the hermitian conjugate fields: 
\begin{eqnarray}
\bar{H}  ^{(Q) a} & = & \gamma ^0 H_a  ^{(Q) \dagger } \gamma ^0 , \nonumber \\
\bar{H}_a  ^{(\bar{Q}) }  & = & \gamma ^0 H  ^{ (Q) \dagger a  }
 \gamma ^0 .
\label{H7}
\end{eqnarray}

The transformation properties of the superfields under the relevant symmetries are summarized in Table \ref{table1}. We deserve special attention to isospin and chiral transformations of $\mathcal{X}$-superfield.
Based on Refs.~\cite{Aceti,XProduction,Gamermann1,Gamermann2,Gamermann3}, we consider that the $X(3872)$ state is generated from the interactions of $(\bar{D}^0 D^{\ast 0} - c.c. )$, $(D^- D^{\ast +} - c.c.)$ and $(D_s ^{-} D^{\ast +} _s - c.c.)$. The motivation comes from the fact that the wave function of $X(3872)$ is very close to the isospin $I=0$ combination of $(\bar{D}^0 D^{\ast 0} - c.c.) $ and $(D^- D^{\ast +} - c.c. )$ and has a sizable fraction of $(D_s ^{-} D^{\ast +} _s - c.c.)$, which tells us that in strong processes the $X(3872)$ behaves as a rather good $I=0$ object and also as $SU(3)_V $ singlet.
Besides, we stress that the nature of $X(3872)$ is still not fully understood; its theoretical interpretation is matter of debate, with various pictures proposed to explain it, as discussed in Introduction and reviews avaliable in literature (see for example Refs. \cite{Brambilla2,Chen}). We can also find older and recent works interpreting $X$ as a charmonium state compatible with the meson $\chi _1 (2P)$ (or a mixture of it and other pictures), since they have similar quantum numbers: $J^{PC}=1^{++}$ \cite{Fazio,Badalian,Ortega,Ferretti1,Ferretti2,Karliner}. We must keep in mind that $\chi_{c1}(2P)$ is a singlet for both isospin and chiral $SU(3)_{L} \times SU(3)_{R}$ transformations. Then, if the $X(3872)$ state is compatible with this radially excited charmonium, they must behave in the same way concerning these transformations. 

Hence, we believe that it seems reasonable to assume $X(3872)$ as an isosinglet and an singlet of $SU(3)_V$. 

\begin{table}
  \caption{ Transformation properties of the superfields above introduced under the relevant symmetries: chiral
$SU(3)_{L} \times SU(3)_{R}$, heavy-quark spin, Lorentz, parity and charge conjugation symmetries. $U$ is a matrix acting on unbroken $SU(3)_{V}$ group;  $S^{(Q)} $ is a rotation matrix acting on heavy-quark spin (HQS); $S^{(\bar{Q})} $ is a rotation matrix acting on heavy-antiquark spin; $D=D(\Lambda)$ is the usual spinor representation of Lorentz transformation $\Lambda$; and $C = i \gamma ^2 \gamma ^0$ is the usual charge conjugation matrix. }
\begin{center}
\begin{tabular}{c|c|c|c|c|c}
%\hline
\hline
Symmetry & $H_a ^{(Q)}$ & $\bar{H}  ^{(Q) a}$  & $H ^{(\bar{Q}) a}$   & $\bar{H}_a  ^{(\bar{Q}) } $ & $\mathcal{X} ^{\mu} $  \\ [4pt]
%\hline
\hline 
 & & & & & \\ [-8pt]
Chiral &  $ H _b ^{(Q)}  U_{ b a }^{\dagger}$  &  $U_{ a b } \bar{H}  ^{(Q) b} $ & $U^{ a b } H ^{(\bar{Q}) b} $  & $ \bar{H}_b  ^{(\bar{Q}) } U_{ b a }^{\dagger} $ & $\mathcal{X} ^{\mu} $   \\ [8pt]
%\hline
HQS & $ S^{(Q)} H_a ^{(Q)} $ &  $ \bar{H}  ^{(Q) a} S^{(Q) \dagger}$ &  $ H^{(\bar{Q}) a} S^{(\bar{Q}) \dagger}$ &  $ S^{(\bar{Q}) } H^{(\bar{Q}) a} $  & $  S^{(Q)} \mathcal{X} ^{\mu} S^{(\bar{Q}) \dagger} $   \\ [8pt]
%\hline
 Lorentz & $ D H_a ^{(Q)} D ^{-1} $ & $ D \bar{H}  ^{(Q) a} D  ^{-1}$ &  $D H  ^{(\bar{Q}) a} D ^{-1}$ &  $D \bar{H}_a  ^{(\bar{Q}) } D ^{-1} $ & $ \Lambda _{\nu} ^{\mu} D \mathcal{X} ^{\nu} D ^{-1} $   \\ [8pt]
 %\hline 
Parity & $ - H_a ^{(Q)} $ & $ - \bar{H}  ^{(Q) a} $ &  $- H  ^{(\bar{Q}) a} $ &  $- \bar{H}_a  ^{(\bar{Q}) } $ & $ - \mathcal{X} _{\mu} $   \\ [8pt]
 %\hline 
Charge C.& $ C H ^{(\bar{Q}) a  \,T} C$ & $ C \bar{H}_a  ^{(\bar{Q})   \,T} C $ &  $C H_a  ^{(Q) \,T} C$ &  $C \bar{H}  ^{(Q) a \,T} C$ & $ - C \mathcal{X} ^{\mu \, T} C = \mathcal{X} ^{\mu} $   \\ [6pt]
 \hline 
\end{tabular}
\end{center}
\label{table1}
\end{table}

Now we are able to introduce the effective Lagrangian describing the interactions among heavy and light mesons and the $\mathcal{X} $-field, respecting the relevant symmetries, i.e. the chiral
$SU(3)_{L} \times SU(3)_{R}$, heavy-quark spin, Lorentz, parity and charge conjugation symmetries. The Lagrangian at lowest order can be written as
\begin{eqnarray}
\mathcal{L} = \mathcal{L}_{M} + \mathcal{L}_{X}.
\label{L1} 
\end{eqnarray} 
The first term in Eq.~(\ref{L1}) carries the kinetic terms and the couplings between light- and heavy-meson fields \cite{Ingsur,Eichten,Georgi,Grinstein,Manohar,Casalbuoni,Abreu,Esposito:2014rxa}:
\begin{eqnarray}
\mathcal{L}_{M} & = & - i \;\mathrm{Tr}\hspace{1pt} \left[ \bar{H}  ^{(Q) b} v 
\cdot \mathcal{D} _{b}^{a}
 \;H_{a}  ^{(Q) } \right]   
 %\nonumber \\ & & 
 - i \; \mathrm{Tr}\hspace{1pt} 
 \left[ H  ^{(\bar{Q}) b}
  v \cdot \mathcal{D} _{b}^{a} \;
 \bar{H} _{a} ^{(\bar{Q})  } \right] \nonumber \\
& & + i g \; \mathrm{Tr}\hspace{1pt} \left[
 \bar{H}  ^{(Q) b} H_{a}  ^{(Q) }
\gamma  ^{\mu} \gamma  ^{5}    \right] (\mathcal{A}_{\mu})_{b}^{a}
 % \nonumber \\ &  &  
 + i g  \;\mathrm{Tr}\hspace{1pt}  \left[H  ^{(\bar{Q}) b}  
\bar{H}_{a}  ^{(\bar{Q})  } 
\gamma  ^{\mu} \gamma  ^{5}  \right] (\mathcal{A}_{\mu})_{b}^{a},
\label{L2} 
\end{eqnarray} 
where
\ben
(\mathcal{D} _{\mu })_{b}^{a}& = & \left[\partial _{\mu} + 
\frac{1}{2} \left( \xi ^{\dagger } 
\partial _{\mu} \xi + 
\xi \partial _{\mu} \xi  ^{\dagger }  \right) \right]_{b}^{a},
 \nonumber \\
(\mathcal{A}_{\mu})_{b}^{a} & = & \frac{1}{2} \left( \xi ^{\dagger } 
\partial _{\mu} \xi - 
\xi \partial _{\mu} \xi  ^{\dagger }  \right)_{b}^{a} ,\nonumber \\
\xi  & = &  e^{\frac{i}{f} M};  \nonumber \\
M & = & \left( \begin{array}{ccc}
\frac{\pi ^{0}}{\sqrt{2}} + \frac{\eta}{\sqrt{6}} & \pi ^+  & K^+ \\
\pi ^- & -\frac{\pi ^{0}}{\sqrt{2}} + \frac{\eta}{\sqrt{6}}  & K^0 \\
K ^-   & \bar{K} ^0                         & - \frac{2 }{\sqrt{6}} \eta
 \end{array} \right).
\label{M} 
\een  
Therefore $M$ represents the light meson fields, with the $\xi$-field transforming as $L \xi U^{\dagger} = U \xi R^{\dagger}$ under chiral $SU(3)_{L} \times SU(3)_{R}$ transformations; $g$ and $f$ are coupling and pion decay constants, respectively.

The last term in Eq.~(\ref{L1}) is the Lagrangian coupling the $\mathcal{X} ^{\mu} $  to heavy mesons: 
\begin{eqnarray}
\mathcal{L}_{X} & = &  \frac{x}{2} \mathrm{Tr}\hspace{1pt}
\left[ \mathcal{X}^{a} _{b\mu} \bar{H}  ^{(Q) b} \gamma  ^{\mu}
 \bar{H}_{a} ^{(\bar{Q}) }  \right]   
% \nonumber \\ & & 
 +  \frac{x}{2} \mathrm{Tr}\hspace{1pt}
\left[ \bar{\mathcal{X}}^{a} _{b\mu} H^{(\bar{Q}) b} (\gamma  ^{\mu})^\dagger
 H_{a} ^{(Q) }  \right], 
  \label{L3}
  \end{eqnarray}
  where $x$ is the coupling constant and $\bar{\mathcal{X}}^{a} _{b\mu} =\gamma ^0\mathcal{X}^{\dagger a} _{b\mu} \gamma ^0$, with $\mathcal{X}^{ a} _{b\mu} $ being diagonal with respect to the light-flavor indices $a$ and $b$. We remark the analogy done here with the situation reported in Ref.~\cite{Esposito:2014rxa}, in which the superfield is related to the $Z_c(3900)$ and $Z_c ^{\prime}(4020)$ states. 

As it can be clearly noticed in Eq.~(\ref{L2}), the coupling between heavy and light mesons are obtained by expanding $H$- and $\xi$-fields and taking the Dirac traces. In particular, the expansion of the axial current $\mathcal{A}_{\mu}$ generates interacions among heavy mesons and an odd number of light mesons. Its leading order is $\mathcal{A}_{\mu} \simeq i \partial _{\mu} M / f$, and therefore $P P^{\ast} M$ and $P^{\ast} P^{\ast} M$ couplings are engendered. Explicitly, these three-body couplings are 
\begin{eqnarray}
\mathcal{L}_{M H H} & = & 
\frac{2 g i}{f} \left[ P ^{\ast (Q)  \dagger b \mu}  P ^{ (Q)} _{ a} - 
 P ^{ (Q) \dagger b} P ^{\ast (Q) \mu} _{ a} \right]\partial _{\mu} M_{b}^{a}
 \nonumber \\
&  & + \frac{2 g i}{f} \; \varepsilon^{\alpha \beta \mu \gamma}P ^{\ast (Q)  \dagger b } _ {\alpha} P ^{\ast (Q)} _{ a \beta } v_{\gamma}
\partial _{\mu} M_{b}^{a} \nonumber \\
& & +\frac{2 g i}{f} \left[ P ^{\ast (\bar{Q}) b \mu} P ^{ (\bar{Q}) \dagger}_{ a} - 
 P ^{ (\bar{Q}) b } P ^{\ast (\bar{Q}) \dagger \mu} _{ a} \right]\partial _{\mu} M_{b}^{a} \nonumber \\
&  & - \frac{2 g i}{f} \; \varepsilon^{\alpha \beta \mu \gamma} P ^{\ast (\bar{Q}) b  } _ {\alpha} P ^{\ast (\bar{Q}) \dagger } _{ a \beta } v_{\gamma}
\partial _{\mu} M_{b}^{a}.
\label{L4} 
\end{eqnarray}

Also, expanding $H$- and $\mathcal{X}$-fields in Eq.~(\ref{L3}) and taking the Dirac traces, we obtain the $P P^{\ast} X$ couplings:
\begin{eqnarray}
\mathcal{L}_{X } & = & 
x \left[ P ^{\ast (Q) \dagger b \mu}  X^{a }_{b \mu } P ^{ (\bar{Q}) \dagger } _{ a} -  P ^{ (Q) \dagger b }  X^{a }_{b \mu } P ^{\ast (\bar{Q}) \dagger \mu} _{ a} \right] \nonumber \\
&  & + x \left[ P ^{\ast (\bar{Q})  b \mu}  X^{\dagger a }_{b \mu } P ^{ (Q) } _{ a} -  P ^{ (\bar{Q})  b }  X^{\dagger a }_{b \mu } P ^{\ast (Q) \mu} _{ a} \right].
\label{L5} 
\end{eqnarray}
It can be remarked the absence of anomalous vertex of type $P^{\ast} \bar{P}^{\ast} X$ in Eq.~(\ref{L5}), i.e. a vertex analogous to the $P^{\ast} \bar{P}^{\ast} M$-term in Eq.~(\ref{L4}). So, this absence (expected due to charge conjugation) is naturally confirmed  in a rigorous way, via explicit calculation of the HMET Lagrangian.

Thus, with the effective Lagrangians obtained above, Eqs. (\ref{L4}) and (\ref{L5}), one can obtain the Feynman rules which will be useful to determine the amplitudes for the relevant processes, which we report in Appendix \ref{AppA}.  

%%%%%%%%%%%%%%%%%%%%%%%%%%%%%%%%%%%%%%%%%%%%%%%%%%%%%%%%%%%%%%%%%%%%%%%%%%%%%%%%%%%%%%%%
%%%%%%%%%%%%%%%%%%%%%%%%%%%%%%%%%%%%%%%%%%%%%%%%%%%%%%%%%%%%%%%%%%%%%%%%%%%%%%%%%%%%%%%%
\section{Scattering amplitudes and cross sections}
\label{Amplitudes} 
%%%%%%%%%%%%%%%%%%%%%%%%%%%%%%%%%%%%%%%%%%%%%%%%%%%%%%%%%%%%%%%%%%%%%%%%%%%%%%%%%%%%%%%%
%%%%%%%%%%%%%%%%%%%%%%%%%%%%%%%%%%%%%%%%%%%%%%%%%%%%%%%%%%%%%%%%%%%%%%%%%%%%%%%%%%%%%%%%
 
Now we can calculate the transition amplitudes associated to the $X$-production in the processes $ \bar{D} D \rightarrow \pi X $, $\bar{D}^* D \rightarrow \pi X $ and $\bar{D}^* D^* \rightarrow \pi X $. 

As discussed in previous Section, we remark that within the present framework the evaluation of $X(3872)$ properties requires its coupling to the components $(\bar{D}^0 D^{\ast 0} - c.c.) $, $(D^- D^{\ast +} - c.c. )$ and $(D_s ^{-} D^{\ast +} _s - c.c.)$. 
 However, due to the nature of the processes considered here, the coupling of $X(3872)$ to the components $(D_s ^{-} D^{\ast +} _s - c.c.)$ does not play role, although they are relevant in reactions analyzed in Refs. \cite{Aceti,XProduction}.      
Accordingly,  the evaluation of amplitudes demands the coupling of $X$ with neutral and charged components of $(\bar{D} D^{\ast } - c.c.)$, which will be performed in the following steps.

\begin{figure}[!h]
    \centering
        \includegraphics[width=12.0cm]{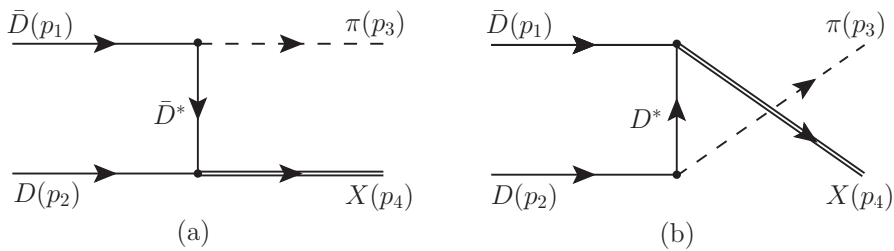}
        \caption{Diagrams contributing to the process $ \bar{D} D \rightarrow \pi X $,  without the specification of the charges.}
    \label{FIG1}
\end{figure}

\begin{figure}[!h]
    \centering
        \includegraphics[width=6.5cm]{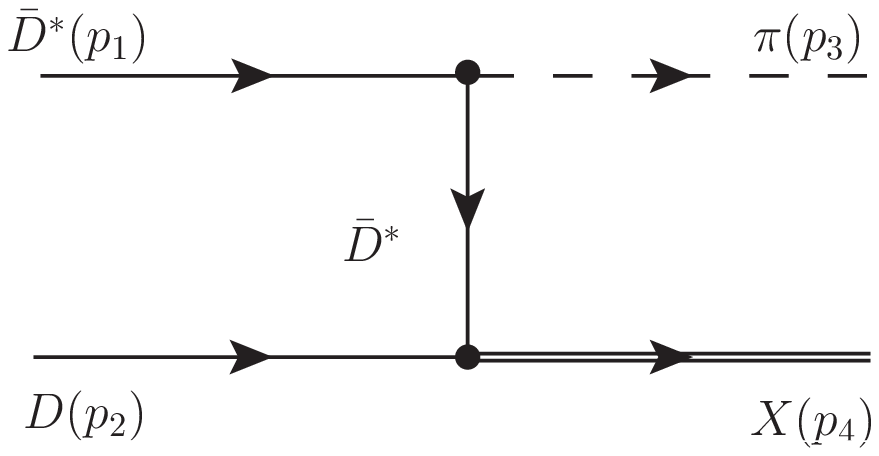}
        \caption{Diagram contributing to the process $\bar{D}^* D \rightarrow \pi X $,  without the specification of the charges.}
    \label{FIG2}
\end{figure}

\begin{figure}[!h]
    \centering
        \includegraphics[width=12.0cm]{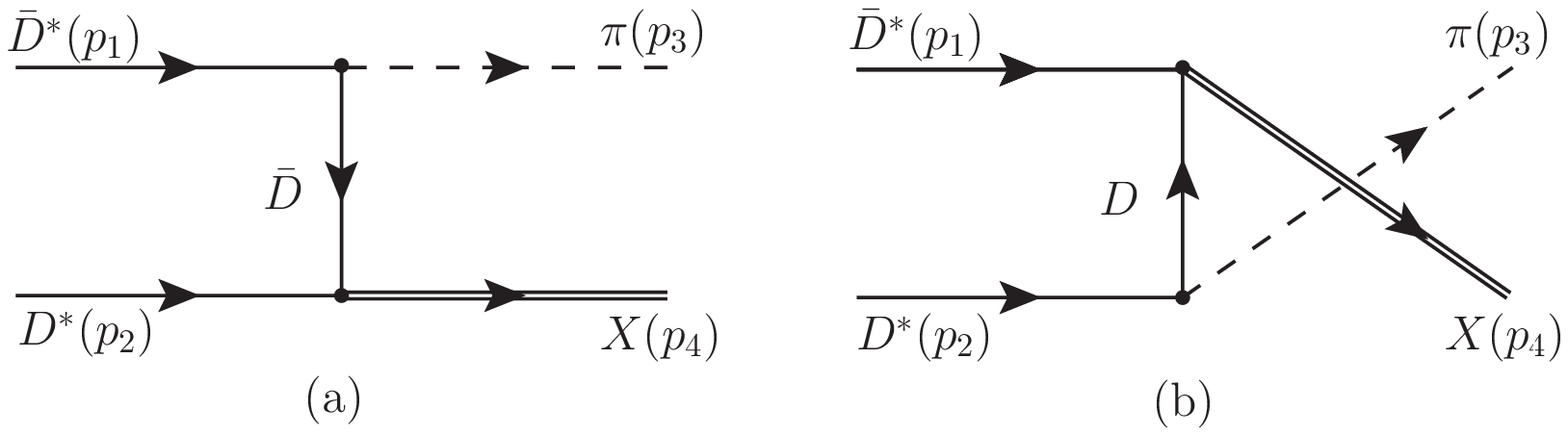}
        \caption{Diagrams contributing to the process  $\bar{D}^* D^* \rightarrow \pi X $,  without the specification of the charges.}
    \label{FIG3}
\end{figure}

In Figs. \ref{FIG1}, \ref{FIG2} and \ref{FIG3} are shown the diagrams contributing to the $ \bar{D} D , \bar{D}^* D , \bar{D}^* D^* \rightarrow \pi X $ processes at leading order, respectively, without the specification of the charges. 

Then, based on the Effective Lagrangians (\ref{L4}) and (\ref{L5}) introduced in previous Section, we find the amplitudes for the diagrams (a) and (b) shown in Fig.~\ref{FIG1} associated to the process $ \bar{D}^{a} (p_1) D^{b} (p_2) \rightarrow \pi^{c} (p_3) X (p_4, \eta) $: 
\ben
 \mathcal{M} _{ 1 } ^{ ( a )} & = &   \frac{1}{2 f} g_{\bar{D}^{\ast d} \pi ^{c} \bar{D}  ^{a} } g_{\bar{D}^{\ast e} D ^{b} X } \delta_{d e} \; \frac{1}{v \cdot k } \;  (p_3 \cdot \eta ^{\ast}), \nonumber \\
\mathcal{M} _{ 1 } ^{ ( b )} & = &  \frac{1}{2 f} g_{D^{\ast d} \pi ^{c} D  ^{b} } g_{D^{\ast e} \bar{D} ^{a} X } \delta_{d e} \; \frac{1}{v \cdot k }\;  (p_3 \cdot \eta ^{\ast}), 
\label{M1}
\een
where $v\cdot k = v \cdot q - m_{D^{\ast}}$, with $q = p_1 - p_3 $ and $q = p_2 - p_3 $ for the diagrams (a) and (b), respectively. The coupling constants have their values according to the charge configuration, and are given in Table \ref{table2}. 

\begin{table}
\caption{ Values of the coupling constants appearing in Eqs. (\ref{M1}), (\ref{M2}) and (\ref{M3}) according to the charge configuration of incoming charmed mesons. } 
\begin{center}
\begin{tabular}{c|c|c}
\hline
\begin{tabular}{@{}c@{}} Charge \\ Configuration \end{tabular} & \begin{tabular}{@{}c@{}} $  g_{D^{\ast d} \pi ^{c} D  ^{a} }$ \\ $( g_{\bar{D}^{\ast d} \pi ^{c} \bar{D}  ^{a} })$ \end{tabular} &  
 \begin{tabular}{@{}c@{}} $g_{D^{\ast e} D ^{b} X }$ \\ $ 
(g_{\bar{D}^{\ast e} \bar{D} ^{b} X })$ \end{tabular} \\
\hline
\multicolumn{3}{c}{1(a)} \\
\hline
$\bar{D}^{0} D^{0}  $ & $ g_{\bar{D}^{\ast 0} \pi ^{0} \bar{D}  ^{0} }\equiv \frac{g}{\sqrt{2}}$  & $g_{\bar{D}^{\ast 0} D ^{0} X } \equiv  \frac{x_n}{\sqrt{2}}$
\\
$D^{-} D^{+}  $ & $ g_{D^{\ast -} \pi ^{0} D  ^{-} } \equiv - \frac{g}{\sqrt{2}}$  & $g_{D^{\ast -} D ^{+} X } \equiv \frac{x_c}{\sqrt{2}} $
\\
$D^{-} D^{0}   $ & $ g_{\bar{D}^{\ast 0} \pi ^{-} D  ^{-} } \equiv g $  & $g_{\bar{D}^{\ast 0} D ^{0} X } \equiv \frac{x_n}{\sqrt{2}} $
\\
 $\bar{D}^{0} D^{+}  $ & $ g_{D^{\ast -} \pi ^{+} \bar{D} ^{0} } \equiv g$  & $g_{D^{\ast -} D ^{+} X } \equiv \frac{x_c}{\sqrt{2}} $
\\
\hline
\multicolumn{3}{c}{1(b)} \\
\hline
 $\bar{D}^{0} D^{0}   $ & $ g_{D^{\ast 0} \pi ^{0} D  ^{0} }\equiv \frac{g}{\sqrt{2}}$  & $g_{D^{\ast 0} \bar{D} ^{0} X } \equiv \frac{x_n}{\sqrt{2}} $
\\
$D^{-} D^{+}$ & $ g_{D^{\ast +} \pi ^{0} D  ^{+} }\equiv -\frac{g}{\sqrt{2}}$  & $g_{D^{\ast +} D ^{-} X } \equiv \frac{x_c}{\sqrt{2}} $
\\
$D^{-} D^{0}    $ & $ g_{D^{\ast +} \pi ^{-} D  ^{0} }\equiv g$  & $g_{D^{\ast +} D ^{-} X }  \equiv \frac{x_c}{\sqrt{2}}$
\\
$\bar{D}^{0} D^{+} $ & $ g_{D^{\ast 0} \pi ^{+} D ^{+} }\equiv g $  & $g_{D^{\ast 0} \bar{D} ^{0} X }  \equiv \frac{x_n}{\sqrt{2}} $
\\
\hline
\multicolumn{3}{c}{2(a)} \\
\hline
  $\bar{D}^{\ast 0} D^{0}  $ & $ g_{\bar{D}^{\ast 0} \pi ^{0} \bar{D}  ^{\ast 0} }\equiv \frac{g}{\sqrt{2}}$  & $g_{\bar{D}^{\ast 0} D ^{0} X } \equiv \frac{x_n}{\sqrt{2}}$
\\
  $D^{\ast -} D^{+} $ & $ g_{D^{\ast -} \pi ^{0} D  ^{\ast -} } \equiv - \frac{g}{\sqrt{2}}$  & $g_{D^{\ast -} D ^{+} X } \equiv \frac{x_c}{\sqrt{2}} $
\\
 $D^{\ast -} D^{0}  $ & $ g_{\bar{D}^{\ast 0} \pi ^{-} D  ^{\ast -} } \equiv g $  & $g_{\bar{D}^{\ast 0} D ^{0} X } \equiv \frac{x_n}{\sqrt{2}}$
\\
 $\bar{D}^{\ast 0} D^{+} $ & $ g_{D^{\ast -} \pi ^{+} \bar{D} ^{\ast 0} } \equiv g$  & $g_{D^{\ast -} D ^{+} X } \equiv \frac{x_c}{\sqrt{2}} $
 \\
 \hline
\multicolumn{3}{c}{3(a)} \\
\hline
$\bar{D}^{\ast 0} D^{\ast 0}  $ & $ g_{\bar{D}^{\ast 0} \pi ^{0} \bar{D}  ^{0} }\equiv \frac{g}{\sqrt{2}}$  & $g_{\bar{D}^{\ast 0} D ^{0} X } \equiv \frac{x_n}{\sqrt{2}}$ 
\\
$D^{\ast -} D^{\ast +} $ & $ g_{D^{\ast -} \pi ^{0} D  ^{-} } \equiv - \frac{g}{\sqrt{2}}$  & $g_{D^{\ast -} D ^{+} X } \equiv \frac{x_c}{\sqrt{2}} $
\\
$D^{\ast -} D^{\ast 0} $ & $ g_{\bar{D}^{\ast 0} \pi ^{-} D  ^{-} } \equiv g $  & $g_{\bar{D}^{\ast 0} D ^{0} X } \equiv \frac{x_n}{\sqrt{2}}$
\\
 $\bar{D}^{\ast 0} D^{\ast +}   $ & $ g_{D^{\ast -} \pi ^{+} \bar{D} ^{0} } \equiv g$  & $g_{D^{\ast -} D ^{+} X } \equiv \frac{x_c}{\sqrt{2}} $
\\
\hline
\multicolumn{3}{c}{3(b)} \\
\hline
 $\bar{D}^{\ast 0} D^{\ast 0} $ & $ g_{D^{\ast 0} \pi ^{0} D  ^{0} }\equiv \frac{g}{\sqrt{2}}$  & $g_{D^{\ast 0} \bar{D} ^{0} X } \equiv \frac{x_n}{\sqrt{2}} $
\\
$D^{\ast -} D^{\ast +} $ & $ g_{D^{\ast +} \pi ^{0} D  ^{+} }\equiv -\frac{g}{\sqrt{2}}$  & $g_{D^{\ast +} D ^{-} X } \equiv \frac{x_c}{\sqrt{2}} $
\\
$D^{\ast -} D^{\ast 0}  $ & $ g_{D^{\ast +} \pi ^{-} D  ^{0} }\equiv g$  & $g_{D^{\ast +} D ^{-} X }  \equiv \frac{x_c}{\sqrt{2}} $
\\
$\bar{D}^{\ast 0} D^{\ast +} $ & $ g_{D^{\ast 0} \pi ^{+} D ^{+} }\equiv g $  & $g_{D^{\ast 0} \bar{D} ^{0} X }  \equiv \frac{x_n}{\sqrt{2}} $
\\
\hline
\end{tabular}
\end{center}
\label{table2}

\end{table}

In the case of the amplitudes for the diagram shown in Fig.~\ref{FIG2} associated to the process $ \bar{D}^{\ast a} (p_1, \epsilon) D^{b} (p_2) \rightarrow \pi^{c} (p_3) X (p_4, \eta) $,  we have
\ben
 \mathcal{M} _{ 2 } & = &   \frac{1}{2 f} g_{\bar{D}^{\ast d} \pi ^{c} \bar{D}  ^{\ast a} } g_{\bar{D}^{\ast e} D ^{b} X } \delta_{d e} \; \frac{1}{v \cdot k }  \;
 \varepsilon^{\alpha \beta \mu \nu} p_{3 \mu} v_{\nu} \epsilon_{\beta} \eta ^{\ast}_{\alpha},
 % \nonumber \\
\label{M2}
\een
where $v\cdot k = v \cdot q - m_{D^{\ast}}$, with $q = p_1 - p_3 $; the coupling constants have their values according to Table \ref{table2}.

%\begin{table}
%\revision{
%\caption{ Values of the coupling constants appearing in Eq. (\ref{M2}), according to the charge configuration.  } 
%\begin{center}
%\begin{tabular}{c|c|c}
%\hline
% Configuration & $  g_{D^{\ast d} \pi ^{c} D  ^{\ast a} }$ &  $g_{D^{\ast e} D ^{b} X }$  \\
%\hline
%  $\bar{D}^{\ast 0} D^{0}  \rightarrow \pi^{0} X  $ & $ g_{\bar{D}^{\ast 0} \pi ^{0} \bar{D}  ^{\ast 0} }\equiv \frac{g}{\sqrt{2}}$  & $g_{\bar{D}^{\ast 0} D ^{0} X } \equiv x_n$
%\\
%  $D^{\ast -} D^{+}  \rightarrow \pi^{0} X  $ & $ g_{D^{\ast -} \pi ^{0} D  ^{\ast -} } \equiv - \frac{g}{\sqrt{2}}$  & $g_{D^{\ast -} D ^{+} X } \equiv x_c $
%\\
% $D^{\ast -} D^{0}  \rightarrow \pi^{-} X  $ & $ g_{\bar{D}^{\ast 0} \pi ^{-} D  ^{\ast -} } \equiv g $  & $g_{\bar{D}^{\ast 0} D ^{0} X } \equiv x_n$
%\\
% $\bar{D}^{\ast 0} D^{+}  \rightarrow \pi^{+} X  $ & $ g_{D^{\ast -} \pi ^{+} \bar{D} ^{\ast 0} } \equiv g$  & $g_{D^{\ast -} D ^{+} X } \equiv x_c $
% \\
% \hline
%\end{tabular}
%\end{center}
%\label{table3}
%}
%\end{table}
%

Finally, the amplitudes for the diagrams shown in Fig.~\ref{FIG3} associated to the process $ \bar{D}^{\ast a} (p_1, \epsilon) D^{\ast b} (p_2, \lambda) \rightarrow \pi^{c} (p_3) X (p_4, \eta) $ are given by 
\ben
 \mathcal{M} _{ 3 } ^{ ( a )} & = &   \frac{1}{2 f} g_{\bar{D}^{\ast a} \pi ^{c} \bar{D}  ^{ d} } g_{\bar{D}^{\ast b} D ^{e} X } \delta_{d e} \; \frac{1}{v \cdot k } \;  (p_3 \cdot \epsilon ) ( \lambda \cdot \eta ^{\ast} ), \nonumber \\
 \mathcal{M} _{ 3 } ^{ ( b )} & = &   \frac{1}{2 f} g_{D^{\ast b} \pi ^{c} D ^{d} } g_{\bar{D}^{\ast a} D ^{e} X } \delta_{d e} \; \frac{1}{v \cdot k } \;  (p_3 \cdot \lambda) ( \epsilon  \cdot \eta ^{\ast} ),
 % \nonumber \\
\label{M3}
\een
where $v\cdot k = v \cdot q - m_{D}$, with $q = p_1 - p_3 $ and $q = p_2 - p_3 $ for the diagrams (a) and (b), respectively. Also, the coupling constants have their values according to Table \ref{table2}. 

It is worthy noticing that the HMET used in the present approach is at leading order in $1/M$ ($ M$  being the mass of the heavy meson), in which $D$ and $D^{\ast}$ are by construction degenerate. However, we take here explicitly the physical (different) masses of charmed mesons; this might be understood as a next-leading order effect in HMET. 

The squared scattering amplitudes, averaged over the spins and isospins of the particles in the initial and final states,  can be written as \cite{XProduction}: 
\ben
 \overline{|\mathcal{M}_i|^2} & = &  \frac{1}{d_1 d_2}  \sum_{Spin} \left[ \left| \mathcal{M}_i ^{(0,0)} \right|^2  + \left| \mathcal{M}_i ^{(-,+)}\right|^2 \right.% \nonumber \\ & & 
\left. + \left| \mathcal{M}_i ^{(-,0)}\right|^2 + \left| \mathcal{M}_i ^{(0,+)}\right|^2 \right],
\label{sumamp}
\een
where $i=1,2,3$ denotes the considered process, the superscript $(Q_1,Q_2)$ the charges of the particles in initial state, and  $d_1$ and $d_2$  the spin and isospin degeneracy factors of the initial particle. 
In addition, in order to make the non-relativistic transition, we must fix the velocity parameter to be $v^{\mu} = (1, \vec{0})$. 
Thus, with this choice and the approximation of the sum over the polarizations
\be
\sum \epsilon ^i \epsilon ^{\ast j } \sim \delta ^{i j }
\ee
in Eqs.~(\ref{M1}), (\ref{M2}) and (\ref{M3}), we obtain the following squared scattering amplitudes: 
\ben
\overline{|\mathcal{M}_1|^2} & = & \frac{1}{4}\frac{g^2}{f^2}
\left[ x_n ^2 + x_c ^2 + x_n x_c \right] |\vec{p}_{\pi}|^2 \frac{1}{\left(\tilde{E}_D -  E_{\pi} - \Delta\right)^2} , \nonumber \\
\overline{|\mathcal{M}_2|^2} & = & \frac{1}{16}\frac{g^2}{f^2}
\left[ x_n ^2 + x_c ^2  \right]  |\vec{p}_{\pi}|^2 \frac{1}{ \left(\tilde{E}_{D^{\ast}} -  E_{\pi} \right)^2 } ,  \nonumber \\
\overline{|\mathcal{M}_3|^2} & = & \frac{g^2}{36 f^2}
\left[ \frac{5}{2}\left(x_n ^2 + x_c ^2\right) + x_n x_c \right] 
%\nonumber \\ & & \times 
|\vec{p}_{\pi}|^2 \frac{1}{\left(\tilde{E}_{D^{\ast}} -  E_{\pi} + \Delta\right)^2} , 
\label{sqamp}
\een
where $\vec{p}_{\pi} \equiv \vec{p}_3$ and $E_{\pi} \equiv E_{3}  = \sqrt{m_{\pi}^2 + |\vec{p}_{\pi}|^2 }$ are the tri-momentum and energy of the pion, $\tilde{E}_{D^{(\ast)}} = p_{D^{(\ast)}} ^2 / 2 m_{D^{(\ast)}} $ is the kinetic energy of incoming particle 1 for each respective reaction and $\Delta = m_{D^{*}} - m_{D}$ the difference between the masses of vector and pseudoscalar charmed mesons.

% Thus, in this scenario the modulus of the momentum of the pion becomes the relevant physical variable~\footnote{It might be interesting to compare at formal level the present amplitudes obtained in this work and those obtained in Ref.~\cite{Mehen}, in which the hadronic decays of $X(3872)$ to $\chi _J$ are discussed in a non-relativistic effective theory. }. 
 
Some comments are necessary concerning the approximations  made above. 
The relevant scales for HMET are the heavy scale $M$ and the physical scale $\Lambda_{\chi} = 4 \pi f_{\pi} \sim 1$ GeV. We notice that $p_{\pi}$ is requested to be much less than $\Lambda_{\chi}$. %, which safely occurs considering $ p_{\pi} \lesssim  200$ MeV. 
Besides, the domain of validity of the present approach engenders in Eqs.~(\ref{M1}), (\ref{M2}) and (\ref{M3}) the limit $k \ll M$. Therefore, the allowed range of pion momentum can also be estimated by imposing $p_{\pi} \ll m_X$ ($m_X$ being the mass of $X(3872)$, the other outgoing particle). Thus, we can reliably use the range $p_{\pi} \lesssim 200$ MeV. On the other hand, in this range the kinetic energy of the incoming charmed mesons are non-negligible with respect to $E_{\pi}$ or $\Delta$. Due to this reason, $\tilde{E}_{D^{(\ast)}}$ is taken into account in the propagators present in Eqs.~(\ref{M1}), (\ref{M2}) and (\ref{M3}). 

%This situation might be compared  to one in Ref.~\cite{Mehen}, in which the hadronic decays of $X(3872)$ to $\chi _J$ are discussed in a non-relativistic effective theory: kinetic energsthe amplitudes
%
%This situation  Ref.~\cite{Mehen}, in which the hadronic decays of $X(3872)$ to $\chi _J$ are discussed in a non-relativistic effective theory.

We remember that the amplitudes shown in Eqs.~(\ref{M1}), (\ref{M2}) and (\ref{M3}) must be multiplied by the factor $ \sqrt{8 m_1 m_2 m_X }$ ($m_1$ and $m_2$ being the masses of heavy fields in initial state) to account for the non-relativistic normalization of the heavy-meson and $\mathcal{X}$ fields \cite{Mehen}. However, we will incorporate this factor in the definition of the couplings $x_n$ and $x_c$, and in next Section their values already take it into account, making them with dimension of $E^{1}$. 

At this point we are able to calculate  the isospin-spin averaged cross sections of the processes discussed above, which in CM frame is given by
\ben
\sigma _i = \frac{1}{64 \pi ^2 (E_1 + E_2)^2} \frac{|\vec{p}_{\pi}|}{|\vec{p}|} \int d\Omega \;\; \overline{|\mathcal{M}_i|^2},
\label{crosssection}
\een
with $i=1,2,3$. The four-vectors associated to the incoming charmed mesons are: $p_1 = (E_1, \vec{p})$,  $p_2 = (E_2, - \vec{p})$; and to outgoing particles are: $p_3 = (E_{\pi}, \vec{p}_{\pi})$ and $p_4 = (E_{X}, - \vec{p}_{\pi})$. The total energy of incoming particles can be approximated to $E_1 + E_2 \approx m_1 + m_2 + E_{CM}$, where $E_{CM} = |\vec{p}|^2/ 2 \mu_{12}$ is the collision energy, with $\mu _{12}$ being the reduced mass of incoming charmed mesons~\cite{Braaten3}. Notice that from conservation of energy the pion momentum can be written as function of collision energy: 
\be
 |\vec{p}_{\pi}| \approx \{ [m_1 + m_2 - m_X + E_{CM}]^2 - m_{\pi} ^2 \}^{\frac{1}{2}}.
\label{mompi}
\ee
Then, using the definitions of quantities in CM frame and Eq.~(\ref{mompi}) in Eq.~(\ref{sqamp}), the cross sections for the three reactions can be given properly as function of $E_{CM}$.

%%%%%%%%%%%%%%%%%%%%%%%%%%%%%%%%%%%%%%%%%%%%%%%%%%%%%%%%%%%%%%%%%%%%%%%%%%%%%%%%%%%%%%%%
\section{Discussion}
\label{Discussion}
%%%%%%%%%%%%%%%%%%%%%%%%%%%%%%%%%%%%%%%%%%%%%%%%%%%%%%%%%%%%%%%%%%%%%%%%%%%%%%%%%%%%%%%%

To pursue our investigation of the $X$-production, now we will analyze the cross sections given in Eq.~(\ref{crosssection}). We use the following values for physical quantities and coupling constants~\cite{PDG,Guo2}: $m_{\pi} = 0.1373$ GeV; $m_{D} = 1.8672$ GeV; $m_{D^{*}} = 2.0086$ GeV; $m_X = 3.8717$ GeV;  $g = 0.6 $; $f = 0.0922$ GeV. 
Focusing on $g_{D^{\ast } D  X }$ coupling constants shown in Table~\ref{table2},
the values considered here are those obtained in Ref.~\cite{Guo2} for original coupling constants coupled to neutral and charged channels with dimensions of $E^{-\frac{1}{2}}$ within HMET approach: $0.35$ GeV${}^{-\frac{1}{2}}$ and $0.32$ GeV${}^{-\frac{1}{2}}$, respectively. Nonetheless, remembering that we have incorporated the factor $ \sqrt{8 m_1 m_2 m_X }$ in the definition of the $x_n$ and $x_c$ couplings, this 
yields the following values with which we will work: $ x_n = 3.772$ GeV; $ x_c = 3.449$ GeV.

Also, taking into account the threshold and the region of validity for the pion momentum ($p_{\pi} \lesssim 200$ MeV), then we can estimate the allowed ranges of validity for collision energy:  $274.6 $ MeV $\leq E_{CM} ^{(1)} \lesssim 390$ MeV, $133.2 $ MeV $\leq E_{CM} ^{(2)} \lesssim 250$ MeV and $ E_{CM} ^{(3)} \lesssim 110$ MeV for each respective reaction. 

\begin{figure}[!h]
    \centering
        \includegraphics[{width=8.0cm}]{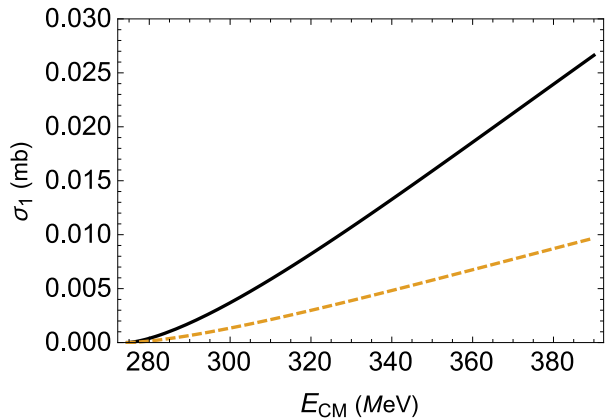} \\
        \caption{Cross section, given by Eq.~(\ref{crosssection}), for the process  $\bar{D} D \rightarrow \pi X $ as function of collision energy. Dashed and Solid lines represent, respectively, the situations considering only the neutral components of $X$ ($x_c = 0$) and with inclusion of charged ones.}
    \label{FIG4}
\end{figure}

\begin{figure}[!h]
    \centering
        \includegraphics[{width=8.0cm}]{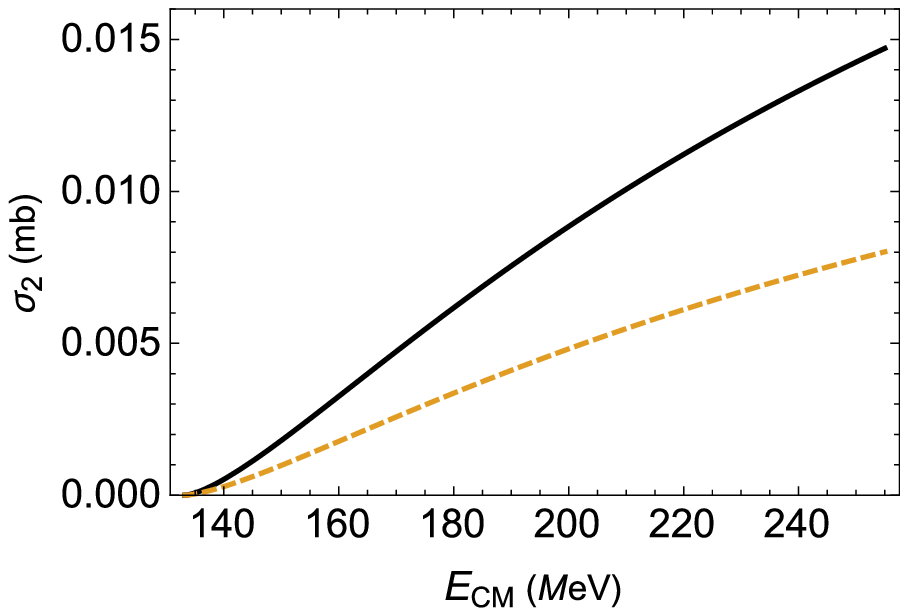} \\
        \caption{Cross section, given by Eq.~(\ref{crosssection}), for the process  $\bar{D}^* D \rightarrow \pi X $ as function of collision energy. Dashed and Solid lines represent, respectively, the situations considering only the neutral components of $X$ ($x_c = 0$) and with inclusion of charged ones.}
    \label{FIG5}
\end{figure}

\begin{figure}[!h]
    \centering
        \includegraphics[{width=8.0cm}]{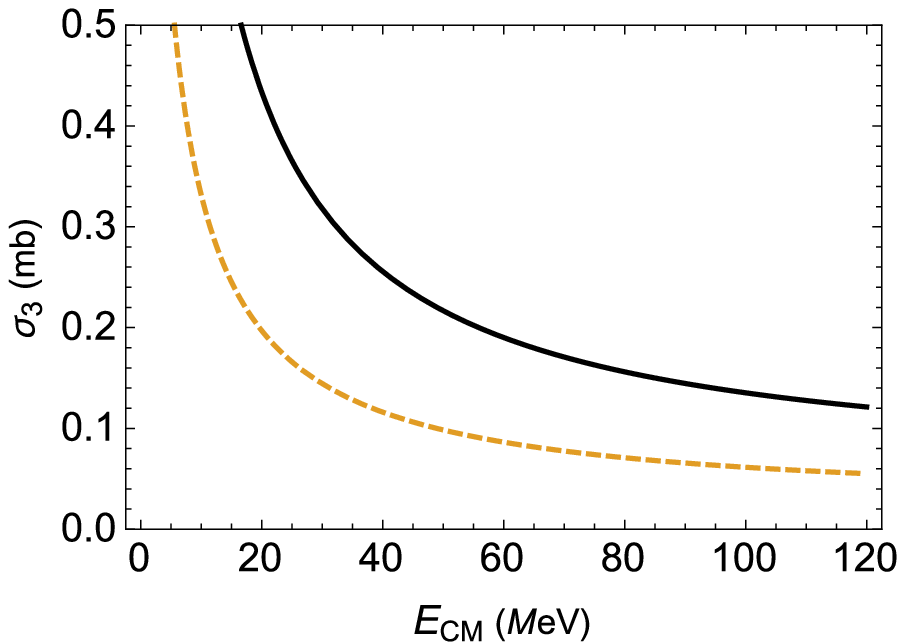} \\
        \caption{Cross section, given by Eq.~(\ref{crosssection}), for the process  $\bar{D}^* D^* \rightarrow \pi X $ as function of collision energy. Dashed and Solid lines represent, respectively, the situations considering only the neutral components of $X$ ($x_c = 0$) and with inclusion of charged ones.}
    \label{FIG6}
\end{figure}

%\begin{figure}[!h]
%    \centering
%        \includegraphics[{height=6.0cm}]{M2.eps}
%        \caption{Squared transition amplitude $\overline{|\mathcal{M}_2|^2}$, given in Eq.~(\ref{sqamp}), as function of pion momentum.}
%    \label{FIG5}
%\end{figure}
%
%\begin{figure}[!h]
%    \centering
%        \includegraphics[{height=6.0cm}]{M3.eps}
%        \caption{Squared transition amplitude $\overline{|\mathcal{M}_3|^2}$, given in Eq.~(\ref{sqamp}), as function of pion momentum.}
%    \label{FIG6}
%\end{figure}

In Figs.~\ref{FIG4}, \ref{FIG5} and \ref{FIG6} are plotted the cross sections in Eq.~(\ref{crosssection}) as function of collision energy $E_{CM}$.
It can be seen that the cross section for the exothermic process, $\bar{D}^* D^* \rightarrow \pi X $, becomes infinity near threshold and acquires the greatest magnitude with respect to the endothermic reactions $ \bar{D} D \rightarrow \pi X $ and $\bar{D}^* D \rightarrow \pi X $ in respective allowed range of $E_{CM}$. In particular, for the pion with momentum $|\vec{p}_{\pi}| \simeq 100 $ MeV, corresponding respectively to $E_{CM} \simeq 307, 166$ and 24 MeV for each reaction, the $\bar{D}^* D^* \rightarrow \pi X $ process yields the biggest cross section by a factor about 70-90 with respect to  other reactions. 

Moreover, the results suggest that the contributions coming from the charged components of the $D$ and $D^*$ mesons enhance the magnitude of amplitudes, which might be relevant in the analysis of production of the state $X(3872)$. Taking again as example the region of $|\vec{p}_{\pi}| \simeq 100 $ MeV, it can be remarked that the charged contributions engender an augmentation of magnitude of cross sections by the factors $\sim 2.75, 1.8 $ and 2.2 in $\sigma _1$, $\sigma _2$ and $\sigma _3$ respectively.

The development reported above can be compared to previous works in more detail.
 We remark Refs.~\cite{ChoLee,XProduction}, which have studied hadronic effects on the $X(3872)$ state in another perspective.

In particular, it can be noticed that in Ref.~\cite{ChoLee} only inverse amplitudes with respect to processes discussed in the present work are reported. In this sense, it is difficult to perform a direct comparison. Nevertheless, we note some crucial differences between mentioned approaches: Ref.~\cite{ChoLee} has been considered only neutral components of $X$; also, the phenomenological Lagrangians furnish $DD^{\ast} \pi$ and  $ DD^{\ast} X$ vertices, which yield only inverse amplitudes with respect to diagrams shown in Figs.~\ref{FIG1} and~\ref{FIG3}. The inverse process of $\bar{D}^* D \rightarrow \pi X $ in Fig.~\ref{FIG3} has not been analyzed, due to the absence of $\pi D^{\ast}D^{\ast}$-coupling. Besides, the coupling constants have different values concerning the ones used here.

On the other hand, a comparison can be done more directly with Ref.~\cite{XProduction}, which is  based on a $SU(4)$ effective approach, and the amplitudes are also constructed by considering both neutral and charged components of $X$. There are essential differences: the first one is obviously the difference of the values of $ D D^{\ast} X$ couplings (we use the couplings obtained via HMET approach). The second distinction is: while in the current work the $ D^{\ast}D^{\ast} \pi$ coupling is the same as $ DD^{\ast} \pi$ coupling, it requires in Ref.~\cite{XProduction} a certain factor due to the nature of $SU(4)$ approach. 

Furthermore, the third dissimilarity is relating to considered vertices: beyond the $ DD^{\ast} \pi$, $ D^{\ast}D^{\ast} \pi$ and $ D D^{\ast} X$ vertices,  it is also used in Ref.~\cite{XProduction} a $ D^{\ast}D^{\ast} X$ coupling, which has been introduced by hand (whose magnitude has been estimated by associate it to the role of triangular loops of mesons), engendering additional diagrams with respect to those in Figs.~\ref{FIG1}, ~\ref{FIG2} and~\ref{FIG3} and giving contributions for the $\bar{D}^* D$ and $\bar{D}^* D^* $ channels. In this scenario (see Erratum of mentioned Ref.) the $\bar{D}^* D^* \rightarrow \pi X $ has the greatest magnitude but with $ \bar{D}^* D \rightarrow \pi X $ being non-negligible, while $\bar{D} D \rightarrow \pi X $ is negligible small, resulting in a different situation of the present case reported above. As debated previously, we emphasize that the expansion of the fields in Eq.~(\ref{L3}) does not yield an interaction Lagrangian involving a $ D^{\ast} D^{\ast} X$ coupling in Eq.~(\ref{L5}).

Notwithstanding, a more realistic comparison can be done by considering the results from Ref.~\cite{XProduction} without $ D^{\ast}D^{\ast} X$ coupling. We see that in this context the findings reported in mentioned Reference tell us that  $ \bar{D} D \rightarrow \pi X $ and $\bar{D}^* D \rightarrow \pi X $ 
reactions are of same order, while $\bar{D}^* D^* \rightarrow \pi X $ is the largest by one-two orders of magnitude. This is qualitatively in agreement with the current work, despite the different energy dependence. 
In addition, all processes in present work have the same order of magnitude with respect to the correspondent reactions in Ref.~\cite{XProduction} without  $ D^{\ast}D^{\ast} X$ coupling. 

%The reason for this difference might be the factor $\sqrt{8}$ in the definition of amplitudes 
%to account for the non-relativistic normalization $ \sqrt{8 m_1 m_2 m_X }$ of the heavy-meson in Eqs.~(\ref{M1}), (\ref{M2}) and (\ref{M3}), as discussed in previous Section. 

Finally, we should also stress another relevant point which distinguishes the present work from the others above mentioned: discussion of region of validity of the framework done, with the results being obtained and analyzed taking it into account.

%Finally, we stress that our work gives a distinct viewpoint to the subject, since it is grounded on HMET, with the effective Lagrangian built respecting chiral $SU(3)_{L} \times SU(3)_{R}$ and heavy quark symmetries. The expansion of the fields in interaction Lagrangians engenders the $ DD^{\ast} \pi$, $ D^{\ast}D^{\ast} \pi$ and $ D D^{\ast} X$
%vertices, allowing a nontrivial amplitude for the process $\bar{D}^* D \rightarrow \pi X $. 
%The contributions related to neutral and charged components are obtained explicitly, as shown in Eq.~(\ref{sqamp}).
%Also, the expansion of the fields in Eq.~(\ref{L3}) does not yield an interaction Lagrangian involving a $ D^{\ast} D^{\ast} X$-anomalous vertex in Eq.~(\ref{L5}), because of violation of charge conjugation.   Besides, the discussion of region of validity of the framework is done in previous Section, and the results can be obtained and analyzed taking it into account.

We conclude this Section with a note on the binding-energy dependence of the couplings $x_n$ and $x_c$. As remarked in Ref.~\cite{Guo2}, when the position of $X(3872)$ approaches the $(\bar{D}^0 D^{\ast 0} - c.c.) $ threshold (and thus $X(3872)$ assumes a long-distance structure), both couplings $x_n$ and $x_c$ goes to zero proportionally to the square root of the binding energy. Thus, at smaller binding energies (i.e. the loosely bound state very near the smallest threshold), the magnitude of cross sections would reduce by approximately the same factor of decreasing of binding energy, since the cross sections depend on linear combinations of quadratic terms involving $x_n$ and $x_c$ couplings.

%%%%%%%%%%%%%%%%%%%%%%%%%%%%%%%%%%%%%%%%%%%%%
%%%%%%%%%%%%%%%%%%%%%%%%%%%%%%%%%%%%%%%%%%%%%
%%%%%%%%%%%%%%%%%%%%%%%%%%%%%%%%%%%%%%%%%%%%%
%%%%%%%%%%%%%%%%%%%%%%%%%%%%%%%%%%%%%%%%%%%%%

\section{Conclusion}
\label{Conclusion}
%%%%%%%%%%%%%%%%%%%%%%%%%%%%%%%%%%%%%%%%%%%%%
%%%%%%%%%%%%%%%%%%%%%%%%%%%%%%%%%%%%%%%%%%%%%
%%%%%%%%%%%%%%%%%%%%%%%%%%%%%%%%%%%%%%%%%%%%%

Summarizing, we have studied the $X(3872)$ production in the processes $ \bar{D} D \rightarrow \pi X $, $\bar{D}^* D \rightarrow \pi X $ and $\bar{D}^* D^* \rightarrow \pi X $, making use of the Heavy-Meson Effective Theory, with the effective Lagrangian built respecting chiral $SU(3)_{L} \times SU(3)_{R}$ and heavy quark symmetries. After the expansion of superfields defined on effective Lagrangian, the amplitudes for these reactions at leading order have been determined, by considering $X(3872)$ as a bound state of $\bar{D}^* D + c.c.$, including neutral and charged components, and obtained the amplitudes of the mentioned reactions. Our findings provide a distinct perspective on this subject, when compared to other works.  A discussion about the region of validity of this framework has been done. Also, we have shown that charged components might play
an important role. We have seen that their contributions increase the magnitude of cross sections by a factor about 1.8-2.75, depending on the reaction.  
besides, it has been shown that in this scenario the $\bar{D}^* D^* \rightarrow \pi X $ has the greatest magnitude concerning the other reactions.

%%%%%%%%%%%%%%%%%%%%%%%%%%%%%%%%%%%%%%%%%%%%%%%%%%%%%%%%%%%%%%%%%%%%%%%%%%%%%%%%%%%%%%%%%%%%%%%%%%%%%%%%%%%%%%%%%%%%%%%%%%%%%%%%%%%%%%%%%%%%%%%%%%%%%%%%%%%%%%%%%%%%%%%%%%%%%%%%
%\begin{acknowledgements}
%%%%%%%%%%%%%%%%%%%%%%%%%%%%%%%%%%%%%%%%%%%%%%%%%%%%%%%%%%%%%%%%%%%%%%%%%%%%%%%%%%%%%%%%

\ack

The author would like to thank the Brazilian funding agency Conselho Nacional de Desenvolvimento Cient\'ifico e Tecnol\'ogico (CNPq) for financial support, through the Grant No. 308890/2014-0.

%\end{acknowledgements}

%%%%%%%%%%%%%%%%%%%%%%%%%%%%%%%%%%%%%%%%%%%%%%%%%%%%%%%%%%%%%%%%%%%%%%%%%%%%%%%%%%%%%%%%
%%%%%%%%%%%%%%%%%%%%%%%%%%%%%%%%%%%%%%%%%%%%%%%%%%%%%%%%%%%%%%%%%%%%%%%%%%%%%%%%%%%%%%%%

%%%%%%%%%%%%%%%
%%%%%%%%%%%%%%%%
%%%%%%%%%%%%%%%%
%%%%%%%%%%%%%%%%
\appendix
%%%%%%%%%%%%%%%%
%%%%%%%%%%%%%%%%
%%%%%%%%%%%%%%%%
%%%%%%%%%%%%%%%%
\section{Appendix: Feynman Rules}

\label{AppA}

In Table \ref{tableApp1} we report the Feynman rules for vertices 
appearing in effective Lagrangians, Eqs. (\ref{L4}) and (\ref{L5}). We consider only the vertices involving pion field. Notice that $x_n$ and $x_c$ mean the coupling constants related to vertices containing neutral and charged heavy mesons, respectively \cite{XProduction}. 

%\begin{largetable}

\begin{table}
 \caption{
 Feynman rules for the vertices 
appearing in the effective Lagrangians, Eqs. (\ref{L4}) and (\ref{L5}). We consider only the vertices involving pion field. In this table, we have used $g_{D^{\ast +} \pi ^{\pm}  D  ^{(\ast) 0} } = g_{D^{\ast 0} \pi ^{\pm }  D  ^{(\ast) +} } = g $, $g_{D^{\ast 0} \pi ^{0}  D  ^{(\ast) 0} } = g/\sqrt{2} $, $g_{D^{\ast +} \pi ^{0}  D  ^{(\ast) +} } = - g/\sqrt{2} $, $g_{D^{\ast -} \pi ^{\pm}  \bar{D}  ^{(\ast) 0} } = g_{\bar{D}^{\ast 0} \pi ^{\pm }  D  ^{(\ast) -} } = g $, $g_{\bar{D}^{\ast 0} \pi ^{0}  \bar{D}  ^{(\ast) 0} } = g/\sqrt{2} $, $g_{D^{\ast -} \pi ^{0}  D  ^{(\ast) -} } = - g/\sqrt{2} $, $g_{\bar{D}^{\ast 0} D ^{0} X } = g_{ D^{\ast 0} \bar{D} ^{0} X }=x_n $, $g_{D^{\ast -} D ^{+} X } = g_{ D^{\ast +} D ^{-} X }=x_c $.}  
\begin{center}
\begin{tabular}{c|c}
\hline
\hline
Process & Rule  \\
\hline
\hline
$D ^{\ast a} (p, \epsilon) \pi ^{b} (q) \rightarrow D ^{c}(p^{\prime} )$  &  $ - \frac{2 i  }{f} \;g_{D^{\ast a} \pi ^{b}D  ^{c} } \;  (q \cdot\epsilon ) $
\\
\hline
$D ^{\ast a} (p, \epsilon) \pi ^b (q) \rightarrow D ^{\ast c} (p^{\prime}, \eta)$  &  \begin{tabular}{@{}c@{}} $ - \frac{2 i  }{f} \;g_{D^{\ast a} \pi ^{b} D  ^{\ast c} }\; $ \\
 $ \times \varepsilon^{\alpha \beta \mu \gamma}   q_{\mu} v_{\gamma} \epsilon _{\alpha}\eta _{\beta} ^{\ast} $\end{tabular}
\\
\hline
$\bar{D} ^{\ast a} (p, \epsilon) \pi^b (q) \rightarrow \bar{D}^{c } (p^{\prime})$  &  $  \frac{2 i   }{f} \;g_{\bar{D}^{\ast a} \pi ^{b} \bar{D}  ^{c} }\;  (q \cdot\epsilon ) $
\\
\hline
$\bar{D} ^{\ast a} (p, \epsilon) \pi^b (q) \rightarrow \bar{D} ^{\ast c} (p^{\prime}, \eta)$  & \begin{tabular}{@{}c@{}} $  \frac{2 i   }{f} \; g_{\bar{D}^{\ast a} \pi ^{b}  \bar{D}  ^{\ast c} }\;$ \\
 $ \times  \varepsilon^{\alpha \beta \mu \gamma}   q_{\mu} v_{\gamma} \epsilon _{\alpha}\eta _{\beta} ^{\ast} $ \end{tabular}
\\
\hline
$D ^{\ast a} (p, \epsilon)  \bar{D} ^{ b} (q) \rightarrow X(p^{\prime}, \eta)$  &  $  - i \; g_{D^{\ast a} \bar{D}  ^{b} X }\;   (\epsilon \cdot \eta ^{\ast}) $
  \\
 \hline 
 $\bar{D} ^{\ast a} (p, \epsilon)  D ^{ b} (q) \rightarrow X(p^{\prime}, \eta)$  &  $  i \; g_{\bar{D}^{\ast a} D ^{b} X }\;  (\epsilon \cdot \eta ^{\ast}) $
  \\
 \hline 
\end{tabular}
\end{center}
\label{tableApp1}
%\end{largetable}
\end{table}

%%%%%%%%%%%%%%%%%%%%%%%%%%%%%%%%%%%%%%%%%%%%%%%%%%%%%%%%%%%%%%%%%%%%%%%%%%%%%%%%%%%%%%%%
%\section*{References}
%%%%%%%%%%%%%%%%%%%%%%%%%%%%%%%%%%%%%%%%%%%%%%%%%%%%%%%%%%%%%%%%%%%%%%%%%%%%%%%%%%%%%%%%

%\References

\end{document}